\documentclass[prl,twocolumn,showpacs]{revtex4}
\usepackage{epsfig}
\usepackage{color}

\newcommand{\be}{\begin{equation}}
\newcommand{\ee}{\end{equation}}
\newcommand{\bea}{\begin{eqnarray}}
\newcommand{\eea}{\end{eqnarray}}
\renewcommand{\r}{{\bf r}}

\begin{document}


\title{ Physical Consequences of Complex Dimensions of Fractals}
\author{Eric~Akkermans$^{1}$\footnote{On leave from Department of Physics, Technion Israel Institute of Technology,
  32000 Haifa, Israel.}, Gerald~V.~Dunne$^{2}$ and Alexander~Teplyaev$^{3}$}
\affiliation{$^{1}$ Department of Applied Physics and Physics, Yale University, New Haven, CT 06520   \\ $^{2}$Department  of Physics, University of Connecticut, Storrs, CT 06269 \\  $^{3}$Department  of Mathematics, University of Connecticut, Storrs, CT 06269}

\begin{abstract}
 It has been realized that fractals may be 
 characterized by complex dimensions, arising from complex poles of the corresponding zeta function, and we show here that these lead to oscillatory behavior in various physical quantities. We identify the physical origin of these complex poles as the exponentially large degeneracy of the iterated eigenvalues of the Laplacian, and discuss applications in quantum mesoscopic systems such as oscillations in the fluctuation $\Sigma^2 (E)$ of the number of levels, as a correction to results obtained in Random Matrix Theory. 
 We present explicit expressions for these oscillations for families of diamond fractals, also studied as hierarchical lattices.
\end{abstract}

\pacs{05.45.Df, 05.60. Gg, 73.23.-b, 68.65.-k}


\maketitle


Fractals, such as the well-known Sierpinski gasket, have been thoroughly studied in physics and in mathematics.
In addition to their own intriguing properties, they provide a useful testing ground to investigate properties of disordered classical or quantum systems \cite{review1},  addressing such fundamental physical issues as Anderson localization, the renormalization group, and phase transitions \cite{gefen}. In addition to condensed matter and statistical physics,  fractals have  been considered in other contexts such as  gravitational systems \cite{polyakov,ambjorn}, and in quantum field theory \cite{hill}. Despite the large amount of work dedicated to the study of the  spectra  of deterministic fractals, 
explicit expressions for spectral functions such as heat kernels or spectral zeta functions, from which many physical quantities can be derived, have remained elusive. It is well known that the heat kernel $Z(t)$  and zeta function $\zeta(s)$ play  central roles in various fields of physics: from mesoscopic physics \cite{am}, to black holes \cite{hawking},  to quantum field theory on curved spaces such as de Sitter and anti De Sitter spaces \cite{birrell}, to the physics of the Casimir effect \cite{elizalde}. This is largely due to their relation to the notion of the partition function in statistical physics \cite{drouffe}, and to the ubiquity of Schwinger's proper-time formalism \cite{schwinger}.

An important step was to identify the leading contribution to Weyl's small time  expansion of $Z(t)$, showing that it is determined by the fractal's spectral dimension $d_s$ \cite{lapidusweyl}, rather than by its fractal (Hausdorff) dimension $d_h$, as initially conjectured.
The fact that fractals are characterized by a set of  more than one dimension, as opposed to standard Euclidean spaces, illustrates the richness and peculiarity of self-similar structures. 
Spectral properties of deterministic fractals have  recently been considered anew in mathematics, and the notion of complex valued fractal dimensions  has been introduced \cite{lapidusweyl,lapidus}, leading to new results for the zeta function \cite{sasha,grabner,bajorin}.  In this Letter we use and extend these results to study the resulting (log-periodic) oscillations in the heat kernel and related physical quantities. We illustrate these ideas with a special class of fractals known as diamond fractals. These diamond fractals permit simple explicit formulas, yet they exhibit properties representative of a wider class, including the Sierpinksi gasket (we discuss this general class in the conclusions). The diamond fractals  also allow us to vary the  spectral dimension, in particular to values less than, greater than, or equal to the critical dimension 2. 

Log-periodic oscillations have a long history in physics: in the theory of phase transitions \cite{phase}, the renormalization group \cite{rg}, Levy flights and fractals \cite{montroll}, and generally in systems with a discrete scaling property \cite{derrida}. Diamond fractals have also been studied in the physics of  hierarchical lattices \cite{kaufman}.

Our main result is the identification and characterization of a new oscillating behavior of $Z(t)$ at small $t$,  which has implications for various physical quantities. Such oscillations do not exist for smooth manifolds, or even for quantum graphs.
We  apply these considerations to the concrete case of quantum mesoscopic systems \cite{am}, and show that the oscillating behavior can be directly observed in spectral quantities such as the fluctuations of the number of energy levels and the Wigner time delay. We also relate the electric conductance $g$, the associated weak localization corrections $\Delta g$, and universal conductance fluctuations $\delta g^2$ to the fractal zeta function. 

\begin{figure}[ht]
\includegraphics[scale=0.6]{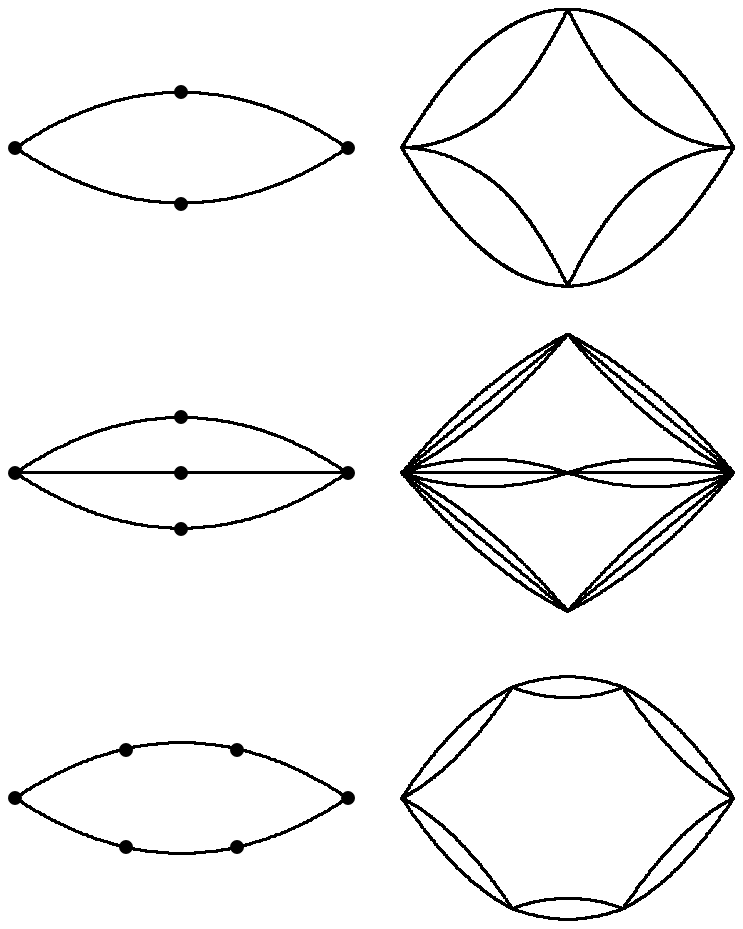}
\caption{First 2 iterations of the diamond fractals $D_{4,2}$, $D_{6,2}$ and $D_{6,3}$.  Their respective branching factors (defined in the text) are $B = 1,2,1$.}
\label{fig3-1}
\end{figure}

We first recall some basic definitions and facts about deterministic fractals. As opposed to Euclidean spaces characterized by translation symmetry, self-similar (fractal) structures possess a dilatation symmetry of their physical properties, each characterized by a specific fractal dimension. To illustrate them, we consider throughout this letter the family of diamond fractals (see Fig. \ref{fig3-1}), but keeping in mind that our results apply to a much broader class of fractals, including the Sierpinski gasket. At each step $n$ of the iteration, we characterize a fractal by its total length $L_n$, the number of sites $N_n$, and the diffusion time $T_n$. Scaling of these  dimensionless quantities allows to define the corresponding Hausdorff $d_h$, spectral $d_s$, and walk $d_w$ dimensions according to
\be
d_h  =  {\ln N_n \over \ln L_n}, \,  \ {d_w} = {\ln T_n \over \ln L_n}, \,  \, \ {d_s} = 2 {\ln N_n \over \ln T_n} \,\, \ ,
\label{dimensions}
\ee
where the limit $n \rightarrow \infty$ is understood. These three dimensions are thus related by $d_s = 2 d_h/ d_w$.

To obtain the heat kernel of a fractal, let us recall the corresponding expression for an Euclidean system of space dimension $d$. We consider the diffusion equation $-\Delta    \psi_k({\bf r})  =E_k \psi_k({\bf r}) $, where the diffusion coefficient is set to unity, without yet specifying boundary conditions. 
The probability $P({\bf r},{\bf r}',t)$ to diffuse, in time $t$,  from an initial point $\r$ to a final point $\r'$,
is given by the Green's function defined in an arbitrary volume $\Omega$:
 $
P({\bf r},{\bf r}',t) = \theta(t) \sum_k \, \sum_{j=1}^{g_k}  \, 
\psi_{k,j}^* ({\bf r}) \psi_{k,j}({\bf r}') \, e^{- E_k t}
$.
Here $g_k$ is a degeneracy factor generally different from unity ({\it e.g.} on a sphere \cite{minak}),
except for one dim. diffusion on a finite interval. 
The heat kernel $Z(t)$ is defined for $t > 0$:
\begin{equation}
\displaystyle Z(t)= \int_\Omega P({\bf r},{\bf r},t) d{\bf r}= 
\sum_k g_k \, e^{- E_k t} \quad .
\label{zdet}
\end{equation}

The spectral zeta function is defined by a Mellin-Laplace transform of the heat kernel
\be
 \zeta (s,\gamma) = {1 \over \Gamma (s)} \int_{0}^{\infty} \frac{dt}{t} \
t^{s} Z(t) e^{- \gamma t} =\sum_{k} \frac{g_k}{(E_k+\gamma)^s} 
\label{zeta1} \ee
Many quantities are derived directly from the spectral zeta function. E.g.,
the spectral determinant $S(\gamma)$ is \cite{hawking}
 \bea
\displaystyle S(\gamma) = \mbox{det}(- \Delta + \gamma ) 
=\exp\left[- {d \over ds} \zeta (s,\gamma) |_{s=0}\right]
\label{detspec1}
\eea
which follows directly from the analytic continuation of $\zeta(s, \gamma)$ in the complex $s$ plane as a meromorphic function analytic at $s=0$, and the  identity $\left[\frac{d}{ds}\lambda^{-s}\right]_{s=0}=-\ln \lambda$. For example, from the spectral determinant, we deduce the density of states: $
\rho(E)=-\frac{1}{\pi}\lim_{\epsilon\to 0^+}{\rm Im}\frac{d}{d\gamma}\ln \, S(\gamma)
$, with $\gamma=-E+i\epsilon$. This can also be written \cite{dashen} in terms of the on-shell S-matrix ${\mathcal S}(E)$ by the Birman-Krein formula $\rho(E)=\frac{1}{\pi}\frac{d}{dE}\ln \,\det\,{\mathcal S}(-E)$, also defining the Wigner time delay: $\tau(E)=-i\hbar \frac{d}{dE}\ln\,\det\, {\mathcal S}(-E)$.

\begin{figure}[hbt]
\begin{center}
\begin{tabular}{||l||r|r|r|r||}
\hline 
& $d_h$ \, \  & $d_w$ \, \ & $d_s = 2 d_h / d_w$ & $l=L_n ^{1/n}$ \\ 
\hline  
 \, \ \ $D_{4,2}$ & $2$ \, \ \   & $2$ \, \ \
& $2$ \, \ \ & $2$ \, \ \  \\
\hline 
 \, \ \  $D_{6,2}$ & ${\ln 6 / \ln 2} $ & $2$ \, \ \ & ${\ln 6 / \ln 2}$ \, & $2$ \, \ \ \\
\hline
 \, \ \  $D_{6,3}$ & ${\ln 6 /  \ln 3} $ & $2$ \, \ \ & ${\ln 6 / \ln 3} $ \, & $3$ \, \ \ 
\\
\hline 
  $\mbox{Sierpinski}$ & ${\ln 3 / \ln 2}$ & ${\ln 5 / \ln 2}$ &
 $2 {\ln 3 / \ln 5}$ \, &
$2$ \, \ \ \\
\hline 
\end{tabular}
\caption{Fractal dimensions and size scaling factor for diamond fractals and for the Sierpinski gasket. For  $D_{6,2}$, the spectral dimension is $d_s\approx 2.58$, and for $D_{6,3}$, $d_s\approx 1.63$.}
\end{center}
\label{table}
\end{figure}
To generalize (\ref{zdet}) to a fractal, we consider the probability $P(r,t)$ to diffuse over a distance $r$ in a time $t$ (with obvious notations). Scaling properties of diffusion are expressed using the definition (\ref{dimensions}) of the walk dimension $d_w$ through the scaling transformation, $P(\lambda r, \lambda^{d_w} t ) = P(r,t)$, for any scaling factor $\lambda$ of the length, so that the probability is of the form $P(r,t) = f (r^{d_w} / t)$, where $f$ is some unknown function. In addition,  the normalization condition, $\int d^{d_h} r P(r,t) =1$, and the change $u = r/ t^{1/ d_w}$, lead to the general scaling form
\be
P(r,t) = {1 \over t^{d_h / d_w} } f(r^{d_w} / t) \quad  .
\ee
This implies that diffusion on a fractal is anomalous in the sense that the usual Euclidean relation $\langle r^2 (t) \rangle \propto  t$, for long enough times, is now replaced by 
$ \langle r^2 (t) \rangle \propto t^{2/d_w} $: hence the name ``anomalous random walk dimension'' for $d_w$. Then, relations (\ref{dimensions}) imply the well known result $P(0,t) \propto t^{- d_s / 2}$ for the leading term of the return probability which is driven by the spectral dimension $d_s$, rather than by the Hausdorff dimension $d_h$. Generalizing (\ref{zdet}), the heat kernel of a diamond fractal can be obtained by noticing that the spectrum of diamond fractals is the union of two sets of eigenvalues. One set is composed of the non degenerate eigenvalues $\pi^2 k^2$, (for $k=1, 2, \dots$).
This corresponds to the spectrum of the diffusion equation defined on a finite one-dimensional interval of unit length, with Dirichlet boundary conditions. The second ensemble contains iterated eigenvalues, $\pi^2 k^2 L_n ^{d_w}$, obtained by rescaling dimensionless length $L_n$ and time $T_n$ at each iteration $n$ according to $L_n ^{d_w} = T_n$, given in (\ref{dimensions}).  To proceed further, we  use the explicit scaling of the length $L_n = l^n$ upon iteration (see Table). These iterated eigenvalues have an exponentially large degeneracy given, at each step, by $B L_n^{d_h} \equiv  B \left( l^{d_h} \right)^n$, where $B =( l ^{d_h -1} -1)$  
is the  branching factor of the fractal (see Fig.\ref{fig3-1}), and the integer $ l^{d_h} $ is the number of links into which a given link is divided.
The {\it exponential} growth of the degeneracy plays a crucial role in our analysis.
By contrast, on an $N$-dimensional sphere the degeneracy  grows as a {\it polynomial}, of order $N-1$ \cite{minak}. 
Finally, the diamond heat kernel $Z_D (t)$ is the sum of  contributions of the two sets of eigenvalues:
\be
Z_D (t) = \sum_{k=1}^\infty e^{- k^2 \pi^2 t} + B \sum_{n=0}^\infty L_n^{d_h} \sum_{k=1}^\infty e^{- k^2 \pi^2 t \, L_n^{ d_w}}  \, . 
\label{hk}
\ee 
The associated zeta function $\zeta_D(s)$, from (\ref{zeta1}) at $\gamma=0$, is
\bea
\zeta_D (s) &=& {\zeta_R(2s) \over \pi^{2s}} \left( 1 + B \sum_{n=0}^\infty L_n^{{d_h}  - d_w s} \right)
 \nonumber \\
&=& {\zeta_R(2s) \over \pi^{2s}} l^{{d_h} -1} \left( {1 - l^{1 - d_w s} \over 1 - l^{{d_h} - d_w s} } \right)\quad ,
\label{zeta3}
\eea
where $\zeta_R(2s)$ is the Riemann zeta function. Note that a very similar structure arises for the Sierpinski gasket \cite{sasha}, with the Riemann zeta function factor replaced by another zeta function.
$\zeta_D (s)$ has complex poles given by
\be
s_m = \frac{{d_h}}{d_w} + {2 i \pi m \over d_w \ln l}=\frac{d_s}{2}+ {2 i \pi m \over d_w \ln l}\quad ,
\label{poles}
\ee
where $m$ is an integer. The origin of these complex poles is clearly the exponential degeneracy factors. The complex poles have been identified with complex dimensions for fractals \cite{lapidus,sasha}.
 
\begin{figure}[ht]
\includegraphics[scale=0.23]{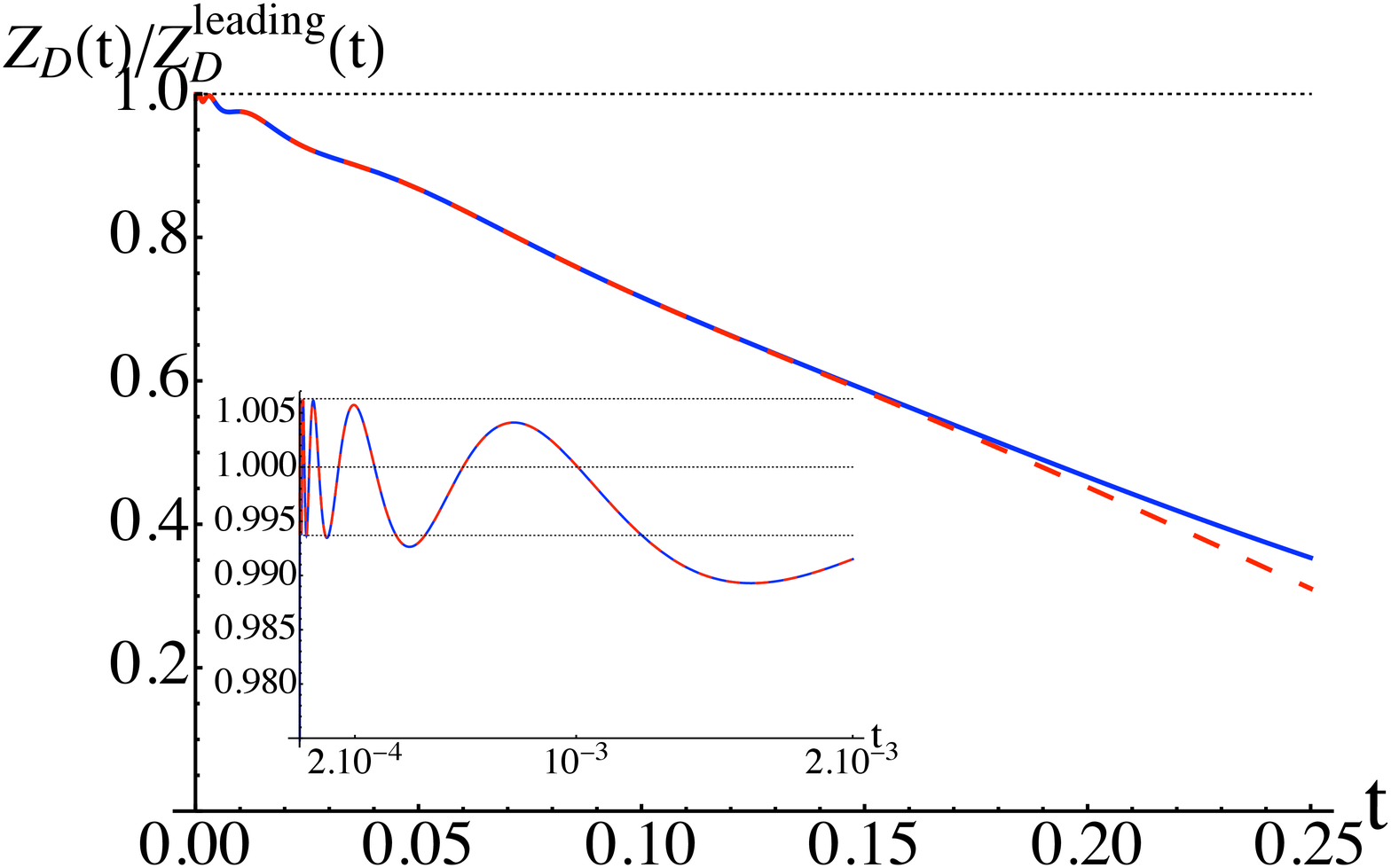}
\caption{Heat kernel $Z_D (t)$ at small time, normalized by the leading non-oscillating term, for the fractal diamond $D_{4,2}$.  The solid [blue] curve  is exact; the dashed [red] curve is the approximate expression  (\ref{zdetosc}). At very small $t$, these curves are indistinguishable, as shown in the inset plot. The relative amplitude of the oscillations  remains constant as $t\to 0$.}
\label{fig2}
\end{figure}

By an inverse Mellin transform, we can write the heat kernel   as $Z_D(t)=\frac{1}{2\pi i}\int_{a-i\infty}^{a+i\infty} ds \,\zeta_D(s) \Gamma(s) \, t^{-s}$. Then the  leading small time behavior comes from the pole of $\zeta_D(s)$ at $s=s_0=d_s/2$, giving the anticipated  time  decreasing function $\sim t^{- d_s /2}$.  The pole of  $\zeta_D(s)$ at $s=1/2$ (coming from the $\zeta_R(2s)$ factor) has zero residue for all diamonds, and so does not contribute to the short time behavior of $Z_D(t)$. (Remarkably, this vanishing of the residue at $s=1/d_w$  also applies to the analogous zeta function on the Sierpinski gasket \cite{sasha}). The pole of $\Gamma(s)$ at $s=0$ gives a constant contribution, $\zeta_D(0)$, to $Z_D(t)$.  But the really surprising new behavior comes from the complex poles in (\ref{poles}), leading to the oscillatory behavior: 
\bea
Z_D (t) &\sim & {l^{d_h -1} -1 \over \ln l^{d_w}} {1 \over t^{d_s /2}} \left( a_0  + 2 \mbox{Re} \left( a_1 t^{- 2 i \pi/(d_w \ln l) } \right)\right) \nonumber\\
&&\hskip 3cm  +\zeta_D(0)+\dots
\label{zdetosc}
\eea
where we have defined $a_m = \Gamma (s_m) \zeta_R(2s_m)/ \pi^{2s_m}$. The leading term $\propto t^{- d_s /2}$ is therefore multiplied by a periodic function of the form $a_{1r} \cos ( \ln t^{s_{1i}} ) + a_{1i} \sin ( \ln t^{s_{1i} })$, where $a_{1r,i}$ are respectively the real and imaginary parts of $a_1$, and $s_{1i} = 2 \pi / \ln l^{d_w}$. The oscillations of $Z_D (t)$ are represented in Fig. \ref{fig2}, and we note that the higher complex poles give much smaller contributions. Similar behavior has been found numerically for the Sierpinksi gasket \cite{strichartz}; from our work, we further find  explicit  expressions for the coefficients, also in the Sierpinksi case.

In principle, all spectral properties can be derived from the heat kernel (\ref{hk}), or from the associated zeta function $\zeta_D (s)$ in (\ref{zeta3}), even though those are not directly accessible physical quantities. For example, the constant term $\zeta_D(0)$ in (\ref{zdetosc}) leads to a topological term $\zeta_D(0)\delta(E)$ in the density of states. More interestingly, the oscillations of $Z_D(t)$ lead to oscillatory behavior in physical quantities. 

\begin{figure}[ht]
\includegraphics[scale=0.23]{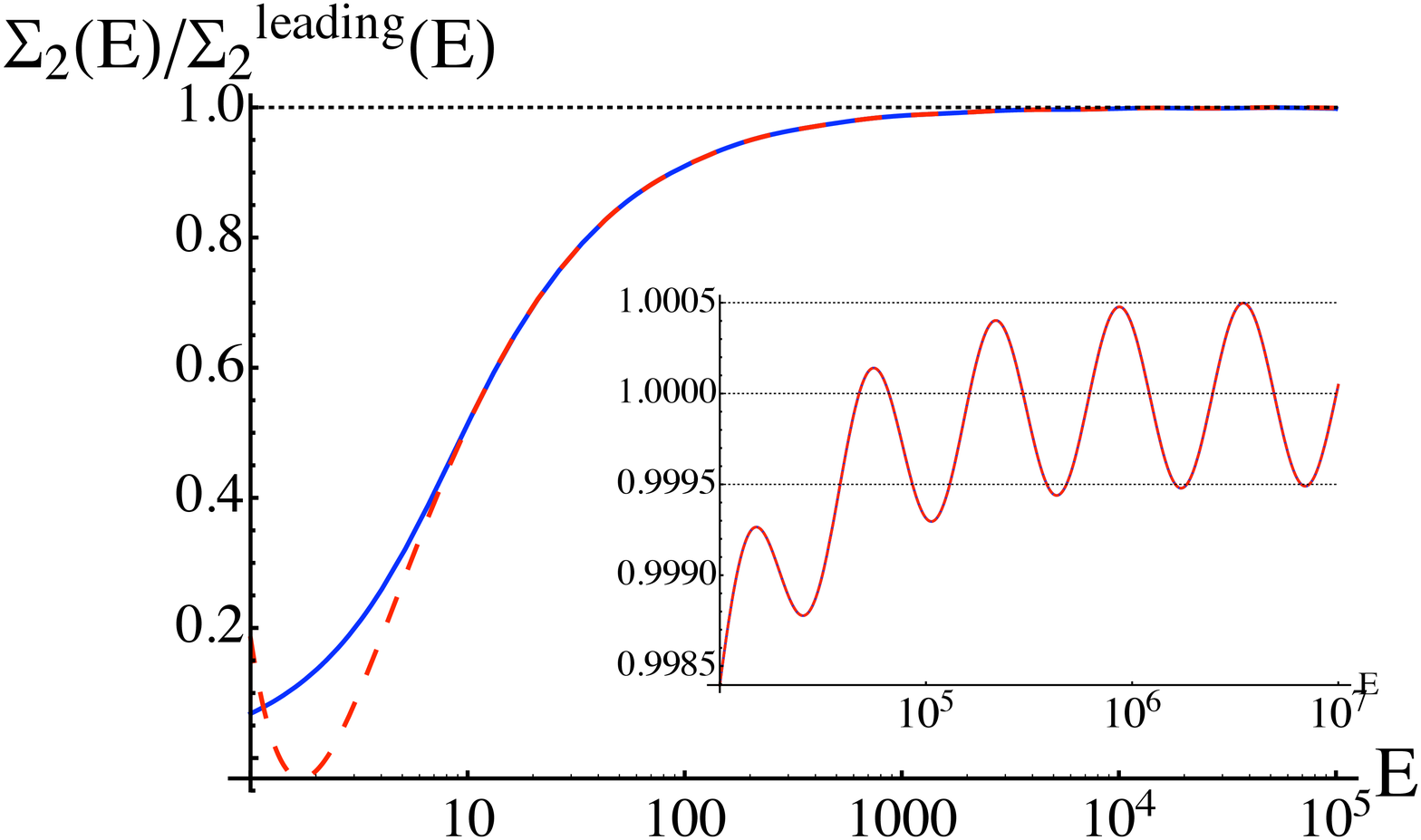}
\caption{Semi-log plot of the fluctuation $\Sigma^2 (E)$, normalized by the leading non-oscillating part,  for the diamond fractal $D_{4,2}$.  The solid [blue] curve is exact; the dashed [red] curve is the approximation  (\ref{sig2D}). At very large $E$, these curves are indistinguishable, as shown in the insert.  The relative amplitude of  oscillations remains constant as $E\to \infty$.}
\label{fig3}
\end{figure}

We  give an explicit example of one  such quantity, in quantum mesoscopic systems. The fluctuation $\Sigma^2 (E)$ of the number of levels within an energy interval of width $E$ is defined by the variance, $ \Sigma^2 (E) = \overline{N^2 (E)} - \overline{ N(E)} ^2 $, of the integrated density of states (the counting function).
In the diffusion approximation, one can express $\Sigma^2 (E)$ directly in terms of the heat kernel through \cite{am}
\be
\Sigma^2 (E) = { 2 \over \pi^2} \int_0^\infty {dt \over t} \, Z_D (t) \, \sin^2 \left( {E t \over 2} \right) \, .
\label{sig2def}
\ee
Inserting (\ref{zdetosc}) for $Z_D (t)$, we obtain,
\begin{eqnarray}
\Sigma^2 (E) &\sim& {l^{d_h -1} -1 \over 2 \pi \ln l^{d_w}} E^{d_s /2} \left( b_0  + 2 \mbox{Re} \left( b_1 E^{ 2 i \pi / \ln l^{d_w} } \right) \right)   \nonumber \\
&&\hskip -1.5cm  -{l^{d_h -1}  \over 2 \pi^2 (l^{d_h} -1)} \left( (l-1) \ln 4E - l\, \frac{l^{d_h -1} -1}{l^{d_h} -1} \ln l^{d_w} \right)
\label{sig2D}
\end{eqnarray}
where  $b_m = \zeta_R (2s_m)/({s_m} \pi^{2 s_m} \sin (\pi s_m ))$. The leading term $\propto E^{d_s /2}$ is now multiplied by a periodic function of the form $b_{1r} \cos ( \ln E^{s_{1i}} ) + b_{1i} \sin ( \ln E^{s_{1i} })$, where $b_{1r,i}$ are respectively the real and imaginary parts of $b_1$. This oscillating behavior of $\Sigma^2 (E)$ is represented in Fig. \ref{fig3}.  It is remarkable that the behavior of $\Sigma^2 (E)$ differs at low energy from the expected ergodic regime independent of fractal dimensions which is well described by Random Matrix Theory, and also at large energy from free diffusion $\propto E^{d_s / 2}$ on the fractal. 

 Transport quantities such as the dimensionless conductance (expressed in units of $e^2 / h$), are also  interesting on a fractal. For instance, the so-called weak localization correction $\Delta g$ to the conductance, and conductance fluctuations described by the variance $\delta g^2$, a universal quantity independent of the system size, take a general and  remarkable form \cite{am} expressed  in terms of the zeta function only, namely  
 $\Delta g = -2 \zeta_D (s=1)$  [N.B. the pole must be treated properly in the special case $d_s=2$], 
  and $ \delta g^2 = 12 \zeta_D (s=2) $. 
For a single channel setup, it is also possible to relate the Fano factor \cite{beenakker} which characterizes shot noise to $\Delta g$ \cite{brouwer} and thus to the zeta function, and we obtain immediately $F = - \Delta g = 1/3$ and $\delta g^2 = 2/15$ for diffusion on a finite one-dimensional interval. On a fractal, and using (\ref{zeta3}), these quantities now depend on the fractal dimensions $d_h$ and $d_s$, and therefore conductance experiments could be used to determine them. 

To summarize, we have considered spectral properties of deterministic fractals such as the heat kernel and the spectral zeta function. Using the class of diamond fractals, we have derived simple and explicit formulas which illustrate a new and general oscillatory behavior of the heat kernel, and  relate it to complex poles, also identified as complex fractal dimensions, resulting from the exponentially large degeneracy of the iterated eigenvalues of the Laplacian. These oscillations which show up in a variety of interesting physical quantities, characterize a fractal. Our results may be useful to study properties of more general quantum graphs \cite{am} where degeneracies must properly be taken into account, and to investigate magnetic \cite{aaas} and topological properties \cite{avron} of fractals when submitted to external fields such as a Aharonov-Bohm fluxes. They may also have interesting implications for gravitational and quantum field theoretic applications.

We end by stating the class of fractals to which our results apply. We have chosen to illustrate our  results using the class of diamond fractals because this class permits simple and explicit formulas. But the general observations about complex dimensions of fractals, and oscillations in the heat kernel trace and associated physical quantities, generalize to the class of fractals known as
finitely ramified self-similar fractals with full symmetry group (i.e.\ the symmetry group has  doubly transitive action on the boundary). A
complete description of these fractals can be found in \cite{shima,malozemov,bajorin}. The best known examples are  the Sierpinski gasket (\cite{strichartz}), the Level-3 Sierpinski gasket and the Vicsek set.  
Note that in these and the majority of other examples the walk dimension is not 2 [as it is for all diamond fractals], and so $d_s$ is not necessarily equal to $d_h$.

Acknowledgements: We acknowledge support from Yale (EA), from the DOE (GD) and from the NSF (AT).

    \end{document}